# 天文物理类英文科技论文写作的常见问题

张双南[1]，许云[2]

（1. 中国科学院高能物理研究所，北京 100049, zhangsn@ihep.ac.cn；
2. 中国科学院上海天文台，上海 200030, xuyun@shao.ac.cn）

2014年2月（第三版）

## 引言

由于文化和语言习惯的问题，中国人写英文文章总会有这样那样的问题，一些不好的中式英语表达习惯会直接影响到英文文章的质量。本文是笔者通过收集十多年来对天文物理类英文期刊论文的审稿、修改工作和自身论文写作，总结出中国学生经常犯的英文错误，以期有助于中国学生提高英文科技论文的写作质量。本文绝大多数例子都是来自于中国学生的英文论文中比较常见的错误或者不流畅的用法，少数例子取自于本文后面所附的参考文献。本文力求实用，从用词分析、标点符号的用法、规范化、语法和句法等方面对常见问题进行简明的分析并给出正确的英文表达，并不追求学院式的理论完整性。

本文的前身，也就是笔者的一篇短文"天文类英文科技论文写作的常见问题"[1]，发表之后，我们收到了很多读者来信和出版社的约稿。本文就是响应读者和出版社的建议而完成的。但是，我们没有接受大部分读者和出版社的建议写一本或者一系列英文科技写作的教材。我们认为，科技英文写作的教材并不特别缺乏，大部分需要用英文写作的学生都上过有关科技英文写作的课。但是从和学生多年的交流和修改学生文章中我们体会到，真正缺乏的是针对中国人的常见英文错误的文献。因此我们决定对原文"天文类英文科技论文写作的常见问题"进行进一步的整理、补充和归纳，并适当地增加我们在其他文献中看到的一些资料和例子，最后形成了本文。

关于本文的篇幅，我们希望尽可能限制在读者能够在一个小时内浏览完全文，并对本文的内容有一些印象，然后再针对自己的英文习惯和可能犯错误的地方回头仔细阅读部分内容，平时英文写作的时候能够很方便地查到自己没有把握的用法和表达。**我们强调不断地"回头阅读"非常重要，因为我们注意到不少学生在声称已经"认真研读"了本文之后，仍然大量重复地犯同样的错误，即使是在导师多次亲笔或者当面多次修改同样的错误之后依然如此。** 这是可以理解的，因为一方面中文的语法和用法相对非常灵活而且容错度很大，另外一方面中文的口语和书面用语差别远远没有英文的口语和书面用语的差别大，而最重要的是中文和英文的各种习惯差别实在是太大太多了，对于长期生活在非英语为母语环境的学生来说，的确很难在很短时间内仅仅通过课堂的学习而不反复练习就能够熟练掌握英文的写作习惯。为了尽可能压缩篇幅，对一些不常见的错误，我们没有列入本文。为了进一步节省篇幅，书中的大部分例句我们都没有翻译成中文，因为我们认为读者在绝大多

的情况下都能够理解英文句子的意思，我们只对容易混淆或者误解的地方给出对应的中文意思 。

中国科学技术大学出版社的杜军和老师认真收集了中国科学技术大学天体物理专业老师和同学们的建议和意见，鼓励并邀请我们完成本文，我们对此深表感谢。我们也特别感谢中国科学院西安光学精密机械研究所的王乐老师对本文写作中的一些英语语法理论方面的建议。最后，我们对于指出我们的短文"天文类英文科技论文写作的常见问题"中的笔误和错误的下列读者表示诚挚的谢意：台湾中央大学的胡嘉玲同学和加拿大McGill大学的王一博士。

**本版本是第三版，我们计划将不定期地继续进行更新。因此我们希望和邀请其他读者向我们指出本文的笔误、错误以及更新的建议，以期以后修订时能够得到改进。**

# 1 用词分析

## 1.1 近义词的用法区别

(1)  "方程"的英文是"equation"，不是"formula"，"formula"的意思是"公式"。

(2)  "question"和"problem"翻译成中文都是"问题"，但是英文的意思完全不同，

用法如"ask a question"，"answer a question"，"formulate a problem"，"solve a problem"等；

[1] I have a **question** about the last page of your presentation.

[2] I have a **problem** about the last page of your presentation.

前者的意思是你对"the last page of your presentation"有不清楚或者不明白的东西要问，后者是指你认为"the last page of your presentation"有错误。

(3)  "too", "also"和"as well"翻译成中文都是"也"和"又"的意思，但是位置和用法有明显的区别：

[1] I have done this quickly **too**. （注意"too"的位置，这里表达的意思是和其他人相比）

[2] I have **also** done this quickly. （注意"also"的位置，这里表达的意思是和其他人相比）

[3] In the mean time, I finished this **as well**. （注意"as well"的位置，这里表达的意思是和自己相比，所以"as well"含有"另外"的意思。）

(4) "few"和"several"用法区分。"several"的意思是"几个"，"few"的意思是"几乎没有"或者"少"，但是"a few"的意思是"几个"。如：

[1] I bought **several** books today. [1] (b) I bought a few books today.

这两句话的意思略微不同：前者情况是指买的比较多，而后者指的是买的比较少。

[2] **Very few** students came to today's seminar. (今天参加讲座的学生很少。)

[3] Few students came last night. (昨晚只来了几个学生。)

[4] **A few** students came last night. (昨晚来了几个学生。)

(5) "property"，"character"和"characteristic用法区分"。它们翻译成中文都是"特征"、"特性"和"性质"，但是在英文中意思不同，很多学生都误用。

[1] "property"是最普通用词，既可指有形或无形的特性，又可指个性或共性的特征。比如：A star has the following **properties**.

[2] "character"多指一类人或事物所具有的独特的典型的特征，用于人的时候比较多。比如：Soldiers should have brave **character**. "character"也常常用于指电影或者书里面的"人物"，比如：David is **a character** in this movie.

[3] "characteristic"指某人或某物天生有别于他人或他物的内部特质或外表特征。比如：The **characteristics** of neutron stars are listed in table 1.

(6) "obscure" (动词和形容词) 的主要意思是"遮挡"。很多同学以为"obscure"的意思就是"模糊"，在应该用"fuzzy"、"blur"、"dark"、"faint"等的时候误用"obscure"。

[1] The black hole is **obscured** by the dust along the line of sight.

[2] The image of the star is **fuzzy (或blurred)**, because the telescope is out of focus.

[3] The source appears to be **dark** (或faint), because of the **obscuration** by the gas in front of it.

(7) "at last"和"finally"不完全一样。如"**Finally**, we reconstruct the expansion history of the universe up to $z$=7 with the distance moduli of SNe Ia and GRBs." "**Finally**" 换成 "**At last**"，意思是不同的：前者的意思是，"最后，我们重建（了）......"，后者的意思是"我们终于重建（了） ......"。

(8) "suspect or suspicious"（正面）和"doubt or doubtful"（负面），翻成中文都是"怀疑"，但是英文的意思几乎是相反的。例子：

[1] We **suspect** that the observed deviation of data from the model prediction is caused by the

over-simplification of the model. (我们**怀疑**观测数据和模型的偏离**是**由于模型过分简化造成的。)

[2] It is **suspicious** that the observed deviation of data from the model prediction is caused by the over-simplification of the model. (观测数据和模型的偏离**有可能是**由于模型过分简化造成的。)

[3] We **doubt** that the observed deviation of data from the model prediction is caused by the over-simplification of the model. (我们**怀疑**观测数据和模型的偏离**不是**由于模型过分简化造成的。)

[4] It is **doubtful** that the observed deviation of data from the model prediction is caused by the over-simplification of the model. (观测数据和模型的偏离**不太可能是**由于模型过分简化造成的。)

(9) 注意区分"distinction" (定性的区别) 和"difference" (定量的不同)：

[1] A **distinction** should be made between a star and a planet. （"star"和"planet"有定性的不同）

[2] The **difference** between the two measurements is only 10%. （显然是定量的不同）

(10) "due to", "because of", "since", "because"和"as"都有"因为"和"由于"的意思, 但是使用起来很不一样。

[1] However, **due to the fact that** (注意此处"the fact that"不能缺少) the inner disk temperature or BB temperature is lower than several keV, the disk or NS flux below 2 keV contributes a considerable portion in the whole energy band.

或者可以改为：

[2] However, **because of** the fact that the inner disk ....

[3] However, **since** the inner disk ....

[4] However, the disk or NS flux below 2 keV contributes a considerable portion in the whole energy band, **because** the inner disk or BB temperature is lower than several keV.

推荐表达[4], 因为主次关系明确, 而且"because"的主要意思就是"因为", 而"since"的主要意思是"自从"(描述时间)。不建议使用"as", 因为"as"的主要意思是"作为"。以下是使用"due to"和"because of"的例子：

[5] **Due to** the expansion of the universe, the distances between galaxies become farther and

farther.

[6] **Because of** your work, this problem has been solved completely.

在大多数情况下，"due to"和"because of"可以通用，但是意思略微有所不同。"due to"指自然发展、顺理成章的理由，"because of"指主动介入导致结果的理由。

(11) "revise", "correct", "modify"的区别：

[1]"revise or revision or revised"意为"修改或者更新，没有错误，但是不够完善"，比如：The new version of the manuscript has been **revised** by updating the reference list.

[2]"correct or correction or corrected"意为"修正或者改正，有错误，需要改过来"，比如：We have **corrected** three mistakes.

[3]"modify"的意思介于两者之间，比如：We have **modified** the introductory part of the paper, by including more background information and references。

(12) "lightly"和"slightly"

　"NELGs are intermediate Seyfert galaxies whose broad line regions are **lightly** (应改为**slightly**) obscured."

(13) "luminous"和"bright"

在天文学中，"luminous"和"bright"的含义是不同的。"luminous"是指"亮度"很大，"bright"是指"流强"很大，"bright"的反义词是"faint"，但是"luminous"没有反义词（指天文学中没有对应的反义词），但是有时候用"weak"来表示和"luminous"相反的意思，或者就简单地用"less luminous"。例句：

[1] Source A is more **luminou**s than source B because source A is located at a much farther distance than source B, although source B is much **brighter** than source A (或者 although source A is much **fainter** than source B).

[2] Source A has a higher **luminosity** than source B because source A is located at a much farther distance than source B, although source B is much **brighter** than source A (或者although source A is much **fainter** than source B).

(14) just 和 only 不分：

[1] This model can **just** explain the data. 这个模型恰好能够解释那些数据（这是对模型的肯定）。

[2] This model can **only** explain the data. 这个模型只能够解释那些数据（这是对模型的否

定）。

(15) 关于 radiative, radiating 和 radiation 作为形容词的用法

中文"辐射过程"、"辐射转移"和"辐射机制"中的"辐射"翻成英文有两种可能"radiative"和"radiation"，但是意思有所不同。比如"radiative transfer"指的是在"转移"的过程中还有"辐射"，所以"转移"是"辐射的"。但是"radiation transfer"指的是给定的"辐射"如何"转移"，并不关心在这个过程是否还有"辐射"，比如计算伽玛射线在介质中被吸收的过程，不关心伽玛射线被吸收掉之后产生的新的辐射以及新的辐射继续会发生什么。另外，不能使用"radiating transfer"、"radiating process"和"radiating mechanism"，因为"过程"、"转移"和"机制"本身都不会辐射。但是"radiating source"是可以的，意思是"正在产生辐射的源"。"radiative source"也是正确的，意思是"会产生辐射的源"。 "radiation source"是一般的用法，比如"An accelerator is also a **radiation** source and it becomes a **radiating** source when it is turned on."是正确的，"A radioactive source is a **radiation** source"和"A radioactive source is a **radiative** source"都是正确的。

最后，副词只有"radiatively"是正确的，比如"The electrons cool down **radiatively**"是正确的表达。

(16) figure, plot, picture, photo, illustration, schematic, diagram, 和 flow chart 的区别

"figure"是统称 "plot"指数据点或者曲线画出的图， "picture"和"photo"指图片或者照片或者图片，"illustration"、"schematic"、"diagram" 都是指示意图，"flow chart"指流程图。

[1] In **Figure** 3, the left panel is the **plot** between the observed temperature and luminosity of this star.

[2] The right panel of **Figure** 3 is the **picture** (或者 photo) of the galaxy. （注意 picture 指一般性的图片，比如手画的，但是 photo 特指照片，一般指"照相机"拍照的）

[3] Figure 4 is an **illustration** of the state transition processes of this binary system.

[4] The **schematic** of this telescope is shown in Figure 7.

[5] The diagram in Figure 3 shows how a star evolves.

[6] Figure 6 is the **flow chart** of our numerical calculations.

(17) limit, limitation, constrain, constraint 都是"限制"的意思，但用法不同

[1] We **limit** （这里"**limit**"的意思是"限制"，不能用"**constrain**"） the redshift range between

0.1 and 0.3 (或者 from 0.1 to 0.3，但是注意不能用 between 0.1 to 0.3，也不能用 between 0.1 to 0.3).

[2] The range of the redshift of these objects is **constrained** （这里"constrain"的意思是"约束"，不能用 limit）between 0.1 and 0.3 with their measured emission lines.

[3] From the measured emission line of this object, we obtain a **constraint** （这里 constraint 是名词，不能用 constrain, 因为 constrain 是动词）to its redshift between 0.1 and 0.3.

[3] The main **limitation** （这里"limitation"的意思是"局限"，不能用 limit 或者 constraint，尽管两者也都是名词）of this method is that it can only be applied to objects at low redshift.

## 1.2 名词的复数形式

(1) 几个单复数同形的常见名词："literature, staff, faculty, deer, sheep，fish，Chinese，Japanese"等。比如：There are one Chinese and two Japanese in this class A.

(2) 易混淆的不可数名词（没有复数形式），如 "evidence"，"equipment"，"horsepower"，"information"，"manpower"等。如果要用复数或者指很多，可以这么用："a lot of equipment"，"95 times of (习惯上"of"也经常省略掉) horsepower"或者"95 horsepower"（这里"horsepower"指的是单位"马力"），"huge manpower"等。

使用"evidence"错误的例子：

[1] There is **one evidence** in support the existence of black holes in the universe.

[2] There are **many evidences** in support the existence of black holes in the universe.

正确的表述：

[3] There is **evidence** in support the existence of black holes in the universe.

[4] There is **one piece of evidence** in support the existence of black holes in the universe.

[5] There are **many pieces of evidence** in support the existence of black holes in the universe.

使用"information"错误的例子：

[6] This observation provides **an information** for the activities of this system.

[7] This observation provides **many informations** for the activities of this system."

使用"information"正确的例子：

[8] This observation provides **information** for the activities of this system.

[9] This observation provides **several pieces of information** for the activities of this system."

[9] This observation provides **a lot of information** for the activities of this system."

(3) 有些抽象名词在具体化时，可以以复数形式出现。表示特指时，可和定冠词连用；表示"某种"或"一次"意义时，可和不定冠词连用[2]。如：How did you smooth away the **difficulties**?（指各种具体困难）It is **a great pleasure** to talk with you. What **a surprise**!

(4) 表 1 中是一些具有不规则复数形式的单数及其复数，注意：由一个词加 man 或 woman 构成的合成词，其复数形式也是 -men 和-women，如 an Englishman，two Englishmen。但 German 不是合成词，故复数形式为 Germans；Bowman 是姓，其复数是 the Bowmans。

表 1 不规则变化[3]

| 单数 | 复数 | 单数 | 复数 |
|---|---|---|---|
| datum | data | nucleus | nuclei |
| spectrum | spectra | focus | foci |
| medium | media | radius | radii |
| nova | **novae** | locus | **loci** |
| formula | **formulae** | torus | **tori** |
| index | **indices** | modulus | **moduli** |
| continuum | continua | | |

(5) 缩写"AGN"中的"N"既可代表"nucleus"（如**an AGN, two AGNs都是正确的**），也可代表nuclei (如**two AGN也是正确的**)

(6) "different"，"various"和大于1的数字（非整数也是如此）后的名词用复数形式。

"In Figure 7, I show the conversion efficiency of different **kind** (应改为**kinds**) of black hole accretion systems."表2列出常用的几种表达。

表2 几种错误单数形式[3]

| 错误表达 | 正确表达 |
|---|---|
| Different node | Different nodes |
| Various method | Various methods |
| Two advantage | Two advantages |
| Fifteen thermocouple | Fifteen thermocouples |
| 1.3 meter | 1.3 meters |

(7) 指示代词的单复数"this, that, these, those"的用法。错误的例子：

"**This (or that)** two pictures are taken ... "应该改为"**These (or those)** two pictures are taken ... "。

(8) 表示由两部分构成的东西，如："glasses"（眼镜），"trousers"，"clothes"等，若表达具体数目，要借助数量词"pair"（对，双），"suit"（套）。比如："a pair of glasses"， "two pairs of trousers"等。

(9) 集体名词，视其所在的语境采用相应动词或代词的单、复数形式[2]。如：

[1] "committee"：The committee **have** discussed all aspects of the case and have not yet reached agreement.

[2] The committee approved the motion unanimously and directed **its** subcommittee to take immediate action.

[3] "majority"：The party's majority **was** small.

[4] Although the complexes are mapped in detail, the majority **are** not accurately dated.

[5] "number"：A large number of problems **were** encountered.

[6] The number of solutions **was** limited.

(10) 非整数情况下的单复数：小于等于1的就是单数，大于1的就是复数。

(11) "单位符号"没有复数形式：1 keV, 2 keV, 1 km, 2 km, 1 kg, 2 km 都是正确的，但是2 keVs, 2 kgs, 2 kms 都是错误的。但是注意"单位"有复数形式：2 meters, 2 grams, 2 ergs 都是正确的。

(12) 最经常漏掉复数的例子：

[1] From **equation** (10) and (11), we can calculate their relations, as shown in **figure** 4 and 5."其中的"equation"和"figure"都应该用复数形式"equations"和"figures"

[2] as follow 改为 as follows 或者 as following

## 1.3 不定代词

在不定代词的使用中要注意指代一致和主谓一致。

(1) "either/neither, each/everyone"等的用法

[1] **Neither** of the members warrants (注意是单数) formation status.

[2] **Everyone wants** (注意是单数) their (注意不能用his或者her) work published quickly.

(2)"none"的用法

"none"代替"no one, not one, no person, no thing"等做主语时，动词采用单数形式；代替"no persons, no things, not any"等做主语时，动词采用复数形式[2]。如：

[1] **None (= not one)** of the telephones is working.

[2] **None (= not any)** are working.

### 1.4 词义的选取

**1.4.1** 科技文章中为了避免歧义，一般情况下使用单词的第一个意思。

(1) "for"的意思包括"为了"、"因为"，但是在科技文章里面只使用"为了"。需要"因为"的时候用"because"或者"since"。再比如"as"的意思包括"作为"、"因为"等，但是在科技文章里面只使用"作为"。

(2) "while"主要有两个意思："当...... 时候"和"然而"。但是科技英文尽可能使用单词的第一个意思，因此建议只使用"while"为"当...... 时候"，当需要使用"然而"的时候用"whereas"。

**1.4.2** 大部分学生都知道"even"有几个意思，但是很多学生不清楚如何使用。举几个例子：

[1] **Even though** (虽然) a scientist does not make a lot of money, I still choose doing research because ...

[2] **Even if** (尽管) this book costs a lot of money, I will still buy it because ...

[3] This book is very good, **even** (甚至) better than the most popular book written by...

**1.4.3** "primary"的意思，很多学生都在使用"primary"的时候误用作"初始"("initial") 或者"初级"("elementary") 或者"非正式"("preliminary")。实际上"primary"的主要意思是"主要"。例子：

[1] The energy spectrum of this source is consisted of two components; the **primary** (主要) component is a blackbody with a temperature of 1 keV and the **secondary** (次要) component is a power-law with a photon index of $-2$.

[2] After analyzing these data, our **preliminary** (初步) conclusion is that this accretion disk has

an inner disk radius of 20 km.

[3] After analyzing these data, our **primary** (主要) conclusion is that this accretion disk has an inner disk radius of 20 km.

## 1.5 冠词

**1.5.1** 可数名词以单数出现的时候前面必须加冠词"a"或者"the"。"a"是不定冠词，没有特指的含意；"the"是定冠词，专门用来放在特指的名词或者名词性短语之前。

(1) 关于买书的例句：

[1] I bought **a** book about Einstein yesterday. (有很多关于爱因斯坦的书，我买了其中的一本，但是从这句话了你无法了解到底是哪一本。)

[1] I bought **the** book about Einstein we discussed in the class yesterday.

[1] I bought **the** book about Einstein's love stories.

[1] I bought **the** book about Einstein. (尽管没有明确指出，但是谈话的双方都知道是哪一本书，或者世界上只有一本关于爱因斯坦的书。)

[2] I bought **three** books about Einstein's general relativity. (随便三本。)

[3] I bought **the three** books about Einstein's general relativity. (只有三本。)

[4] I bought **the three most** popular books about Einstein's general relativity. (三本最受欢迎的。)

[5] I bought **three most** popular books about Einstein's general relativity. (最受欢迎的其中三本。)

通过比较这几个例句，可知[2]—[5] 的意思完全不同。

(2) 最经常漏掉"the"的例子：

[1] "In left panel of Figure 2"改为"In **the** left panel of Figure 2"；（left panel 是特指）

[2] "In following section, we show…"改为"In **the** following section, we show…"；（following section 是特指）

[3] "We will calculate this quantity in following paragraph."改为"We will calculate this quantity in **the** following paragraph."（following paragraph 是特指）

[4] "In above calculations, we have ... "改为"In **the** above calculations, we have ... "（"above

calculations"是特指。)

[5] From above discussions 改为 From **the** above discussions，道理同上

(3) 习惯用法，same前一定要加"the"

the same as 经常误写为 same with

[1] Our calculation result is **the same as** (注意这里常常犯两个错误，一是忘记加"the"，二是把"as"误写为"with") that reported in Zhang et al. (1989).

the same 误写为 a same

[2] This is **the** (经常忘记加"the"，或者把"the"写成"a") same phenomenon.

[3] The twins look **the** same. 那对双胞胎看上去一模一样。

[4] 比如"**The** same method is also used here."很多同学都错误地写成"Same method is also used here."

[5] 再比如"We have chosen the cross-correlation method, **the** same as that used by Ling & Zhang (2006)."是正确的。

**1.5.2** "single"后面如果是可数名词单数的话，仍然需要在"single"前面加"a"或者"the"，比如"a single book"，"the single tree"。

**1.5.3** "If there is sufficient amount of matter around a black hole ..."修改为"If there is **a** （注意这里需要"a"）sufficient amount of matter around a black hole ..."。

**1.5.4** 不该用但是误用"the"的情况

(1) "the equation 3","the table 4","the figure 5"都是错误的，因为"equation 3","table 4","figure 5"已经是特指了，不需要再加"the"。

最经常误用"the"的例子：

This is shown in **the** equation (10) and the figure 5.

As **the** description above 改为 As described above

as **the** follows 改为as follows （上述例子中的"**the**"都需要去掉）

(2) 物理量符号前面一般不需要加"the"，因为已经是特指了。例如"Empirically, the (需要删

除**the**) γ is roughly correlated to observed peak (需改为**the observed peak**) spectral energy *E*p by lg *E*p =(2.76 ±0.07) － (3.61 ±0.26) lg γ.

We employ this empirical relation to estimate **the** (需要删除**the**) *E*p of Swift GRBs and correct the observed peak luminosity to a bolometric band."

但是"We employ this empirical relation to estimate **the** peak energy *E*p of Swift GRBs and correct the observed peak luminosity to a bolometric band."就是正确的，因为"peak energy"前的"**the**"不是指的"*E*p"，而且指的"peak energy"。

(3) "The role of MXBs in **the** cosmology has attracted increasing attention in recent years."改为"The role of MXBs in cosmology has attracted increasing attention in recent years.", 因为"cosmology"在这里是抽象名词，而如果使用"**the** cosmology"就意味着讨论的有几种"cosmologies", "the cosmology"特指其中的一种"cosmology"。

**1.5.5** 对于英文单词前不定冠词"a"和"an"的用法。元音字面开始的单词前面加"an", 辅音字母开始的单词前面加"a"。

有几个地方需要注意："**an** hour, (因为这里"h"不发音), "a histology class, a one-way path, **a** uniform look" (因为h, o, u 的发音是辅音)。

对于缩写词，有两类用法原则上都是正确的。

(1) 按照缩写的第一个字母代表的单词前面加"a"或者"an"。如"**a** NLS1 galaxy" ("a Narrow Line Seyfert 1 galaxy"), or "**an** UV peak" (an ultraviolet peak) 都是正确的, 。

(2) 按照缩写的第一个字母的发音，元音发音的前面加"an", 辅音发音的前面加"a"。因此"A、E、F、H、I、L、M、N、O、R、S、X"前面加"an", "B、C、D、G、J、K、P、Q、T、U、V、W、Y、Z"前面加"a"。

例如"**a** NLS1 galaxy"和"**a** UV bump"也都是正确的。但是第二种用法更加普遍被接受，因此建议全部使用第二种用法。如果你坚持一定要使用第一种用法，则最好整篇文章保持一致。由于很多英文为母语的人甚至编辑也不理解上面两个用法的道理，有时候即使你整篇文章都采用了第二种用法，也会遇到编辑把"a UV peak"改成"an UV peak"。

**1.6 介词的误用**

**1.6.1** "as"和"in"的误用

The results are shown **as** (应该改为"**in"**) Figure 2.

**1.6.2** "similar to"，"different from"和"differ from"经常被误用为"similar with"或者"similar as"，"different with"和"differ with"。

(1) "Our solution differs **with** (应改为**"from"**) the solutions obtained in previous studies"。或者 "Our solution are different **with** (应改为**"from"**) your solutions"。

(2) 但是下面几个例子是正确的：

[1] We **differ with you** on solving this problem, by including the effects of magnetic fields.

[2] The brothers **differ in** their interests.

[3] The houses in the row **differ** only in small details.

[4] We **differ** about moral standards.

**1.6.3** "to do something"和"in order to do something"是有明显区别的，大部分同学使用时不进行区分。比如：

[1] I have solved this problem **to get** this result. (我已经解决了这个问题，导致我得到了这个结果。)

[2] I have solved this problem **in order to get** this result. (为了得到这个结果，我已经解决了这个问题。)

"**To get** this result I have solved this problem."和"**In order to get** this result I have solved this problem."的意思一样，但是在这种情况下，建议使用"In order to ... "，因为没有引起歧义的可能性。

### 1.7 比较级

(1) 多音节的形容词和副词的比较级/最高级用"more/most+原词"来表示。比如： This problem can be solved using this method **easier**（应该改为"more easily"）.造成上述错误的原因是："easy"的比较级是"easier"，所以很多学生就把"easily"的比较级也错误地认为是"easier"。但是由于"easily"是多音节词，所以没有比较级，只能加"more"、

(2) 注意"better, best, most, least, worse, worst"等取不规则变化的比较级的用法，如

[1] You can do this **better**.

[2] You **better** do this.

[3] This problem is **best** solved by...

[4] This development is **most** advanced.

[5] That development is **least** advanced.

下面两个例句说明fast的比较级和最高级的习惯用法：

[6] He speaks **faster**（不能用"**more fastly**"）than Mary.

[7] He runs **fastest**（不能用"**most fastly**"）in Class 1.

## 1.8 中英文翻译和习惯用法

**1.8.1** 中国地名的英译。

(1) 由专名和通名构成的地名，原则上分写；汉语中单音节的专名一般与通名连拼后再加英文通名。

表3 常见地名翻译[4]

| 太行山脉 | Taihang Mountains |
|---|---|
| 洪泽湖 | Hongze Lake |
| 恒山 | Hengshan Mountain |
| 淮河 | the Huaihe River |
| 巢湖 | the Chaohu Lake |
| 渤海 | the Bohai Sea |

(2) 地名的符号不能省略

a，o，e开头的音节连接在其他音节后面的时候，如果音节的界限易生混淆，用隔音符号，地名中的隔音符号不能省略。如： （陕西）西安市Xi'an City （如果省略隔音符号，就成为Xian，可以读成仙、先、 现、限、鲜、险、县等）

(3) 地名中数字的书写

地名中的数字一般用拼音表示，但数字代码和街巷名称中的序数词用阿拉伯数字表示，见表4。

表4 带数字的地名翻译[4]

| 五指山 | Wuzhi Mountain |
|---|---|
| 九龙江 | Jiulong River |
| 三门峡 | Sanmen Gorge |
| 332高地 | Height 332 |
| 南丹路80号 | 80 Nandan Road |

(4) 专名的英译

陕西省 Shaanxi Province；山西省 Shanxi Province

1.8.2 中式翻译

Larger (smaller)是指"几何"大（小），higher (lower) 是指"性质"大（小）或者高（低），例如

Larger (smaller) flux 改为 higher (lower) flux

Larger (smaller) luminosity 改为 higher (lower) luminosity

Larger (smaller) accretion rate 改为 higher (lower) accretion rate

Larger (smaller) temperature 改为 higher (lower) temperature

Larger (smaller) speed 改为 higher (lower) speed

但是"This box has a larger size than the other one." 是正确的。

表5 几个常见的错误名词中式翻译

| | 正确翻译 | 错误翻译 |
|---|---|---|
| $x^2$ | chi-squared | chi square |
| 平方根 | square root | square-root |
| 均方根 | root mean square (RMS) | root-mean square |
| 长轴 | major axis | long axis |
| 短轴 | minor axis | short axis |
| 上限 | upper limit | up limit |
| 下限 | lower limit | low limit |
| 大质量 | high mass | large mass |
| 小质量 | low mass | small mass |
| 最右边 | rightmost | most right |
| 最左边 | leftmost | most left |
| 最上面 | uppermost | most upper |

| 最下面 | lowermost | most lower |
|---|---|---|
| 多次观测的数据 | multiple observations data | multi-times observations data |
| | multiply observed data | multi-timely observed data |
| 四次观测的数据 | the data observed for four | the four times observed data |
| 视线 | line of sight (LOS) | sight line |

(1) "亮度可以从以下方程计算得到"经常被写成"The luminosity can be calculated from the **below** equation."正确的写法是：

[1] The luminosity can be calculated from the **following** equation.

[2] The luminosity can be calculated from the equation **as follows**. ("**following**"也可)

若将其中的"**following**"换成"**followings**"则是错误的。

(2) 英文习惯中用"for why"连接 explanation, evidence, overview 等先行词。
"Physicists were looking for a fundamental explanation **why** (应改为**for why**) the electron mass could not be any different from its measured value."

(3) "数量级"是"order of magnitude"，但是经常被误用为"order"，例如"The mass of this black hole is **about one order smaller** (应改为"**smaller by about one order of magnitude**") than that one."

(5) "An extra constrained BPL is needed to fit the hard excess above 15 keV due to **small** levels of comptonization at lower luminosity."其中"small"的用法是不对的，因为"small"是指几何的"小"，所以可以改为
[1] An extra constrained BPL is needed to fit the hard excess above 15 keV due to **low** levels of comptonization at lower luminosity. 或者
[2] An extra constrained BPL is needed to fit the hard excess above 15 keV due to **a weak comptonization component** at lower luminosity.

## 1.9 同一单词不同词性的选取

**1.9.1** 同一单词的名词、形容词和动词

(1) "sun"是名词，"solar"是形容词。正确表达："the mass of the Sun"(太阳的质量)，"solar mass"(太阳质量，是质量单位)。"sun mass", "sun energy"等是错误的。

(2) "emission"(可数名词) 的意思是"发射" (动词是"emit"，没有形容词，所以只能使用名词或者动名词作为形容词，如"emission process" （不使用"emitting proce）s"), "emitting source" 和"emission source"都是正确的（发射源）);

"irradiation "(不可数名词) 的意思是"照射"，导致的主要结果是"辐射"会进入到被照射

的对象；

"illumination" (不可数名词) 的意思是"照亮"，导致的主要结果是被照亮的物体表面变亮。

(3) "复杂"的动词、名词和形容词分别是："complicate", "complication"和"complicated" (与"complex"和"sophisticated"大致同义，但是这3个词的含义略有不同："complicated"或者"complication"是指比较乱、多个因素同时并交互起作用；"complex"是指规模比较大，但是并不乱；"sophisticated"指逻辑关系比较复杂)。

(4) "double"是形容词和动词，"twice"是名词。下面是正确的例句：

[1] They discovered a **double** neutron star system.

[2] The **double** neutron stars in the systems have been observed many times.

[3] This observation program has **doubled** the number of neutron stars known so far.

[4] He concluded that QPOs at both the NS spin frequency and **twice** of it should be detected during long (super) burst events for 4U 1636-53.

但是注意"triple"和"quadruple"同时是形容词、名词和动词。

(5) present 的几种用法： At present; is present, At the present time; we present, it is presented,

[1] Our result is **presented** (注意常常被误写成主动式"**present**") in section 3. 或者 "We **present** our result in section 3." 也是正确的。

[2] Two flares **are presented** (注意常常被误写成 "**present**"，误用了动词的主动式) in the light curve of this object.

[3] **At present** (指目前这个时期), we are concentrating on studying the outburst mechanism of this object.

[4] **At the present time** (指目前这个阶段), we are testing this code.

**1.9.2** 形容词和副词、名词和动名词的选取

(1) "relative"和"relatively"。下面一段话中有很多错误： This is mainly caused by the **relative** (改为**relatively**) large statistical **scatters of** (改为**scattering in**) the GRB relations and the **relative** (改为**relatively**) small **data** (改为**number**) of GRBs (此处添加**in the sample,**) **comparing** (改为**compared**) to that of SNe Ia **now** (改为**currently**)."上面的"enough significant"还可以改为"sufficiently significant"。

(2) "evenly"和"even"的区别："All girls should sit on the chairs with **even** (偶) numbers."

"Students are **evenly** (对半) divided into two groups."

(3) the anti-correlation shows larger **scattering**"中的"**scattering**" 改为"scatter".

(4) "study"是动词也是名词，"in this study"常常被错误地写成"in this studying"。

1.9.3 "where"用作关系副词,引导定语从句时，前面经常错误地加"in"，"at"或者"on"。

由于它本身在意义上相当于"介词+which"，所以其前通常无需再用介词。

[1] The black hole is located near the center of NGC 3341, **in** (需要把"**in**"去掉) where many neutron stars move at high speed.

[2] We finally climbed up to the top of the mountain, **on** (需要把"**on**"去掉) where there are many beautiful flowers.

[3] There is a new object at the lower-left corner of Figure 3, **at** (需要把"**at**"去掉) where no source was found before.

但"where"也在句子中充当代词，比如下面的用法是正确的：

[4] Where do you come from？

[5] Where are you going to？

## 1.10    动词的现在分词和过去分词作状语，取决于主句的主语

(1)    "comparing"和"compared"容易误用

**Compared** to the joint constraints with GRBs and without GRBs, we can find the contribution of GRBs to the joint cosmological constraints**,** although the contribution of GRBs to the cosmological constraints would not be significant enough, **comparing** to that of SNe Ia at present.

改为

**Comparing** to the joint constraints with GRBs and without GRBs, we can find the contribution of GRBs to the joint cosmological constraints**,** although the contribution of GRBs to the cosmological constraints would not be significant enough, **compared** to that of SNe Ia at present.

[1] "**Compared**"改为"**Comparing**"，因为该句主语是"**we**"。

[2] "**comparing**"改为"**compared**"，因为"the contribution of GRBs"被和"that of SNe Ia at present"做比较。

再给一个常见的错误例子

**Comparing** (改为**compared**，因为该局的主语是"stellar mass black holes"，只能使用被动语态) to neutron stars, stellar mass black holes are found to be much heavier.

(2) "base","basing"，和"based"容易误用

[1] **Base** (改为"**Basing**") on these data, we can calculate the mass of the black hole.

该句主语是**we**。

[2] **Base** (改为"**Based**") on these data, the mass of the black hole can be calculated.

该句主语是"**the mass of the black hole**"，请注意其中的"on"是必需的。

(3) "accompanied"和"accompanying": "the fall-back matter after the **accompanied** (改为 "**accompanying**") supernova explosion forms an accretion disk around the black hole."或者改为 "the fall-back matter **accompanied by** a supernova explosion forms an accretion disk around the black hole."

(4) "surrounded"和"surrounding": "The accretion disk **surrounded** (改为"**surrounding**") the black hole is made of mostly ionized gas."

"The black hole **surrounding** (改为"**surrounded**") by an accretion disk is growing very fast."

(5) "originated"和"originating", "producing"和"produced"

"The neutron star **originated** (改为"**originating**") from a supernova explosion is about 1000 years old."或者改为"The neutron star produced **by** (或者用"**from**", "**in**"都可以) a supernova explosion is about 1000 years old." 或者改为"The supernova explosion **producing** the neutron star happened about 1000 years old ago."

如果想强调是"过去"产生的，可以这么写"The neutron star **that originated** from a supernova explosion is about 1000 years old."或者"The neutron star that was produced by (或者from，或者 in) a supernova explosion is about 1000 years old." 或者 "The supernova explosion that produced the neutron star happened about 1000 years ago."

(6) Dominating 和 dominated

[1] "这个系统的主导过程是…"：The **dominating** (也可以换成 dominant，但是不能用 dominated) process in this system is…如果写成"The **dominated** (也可以换成 dominant，但是 不能用 dominating) process in this system is…"意思就成了"这个系统中被主导的过程是…"， 意思完全不同了。

[2] Here we study the quantum effects **dominated** systems. (注意不能写成"dominating"，因为 这里指的是研究"systems"，而"systems"是被"quantum effects"主导) 如果写成"Here we study the dominating quantum effects in these systems."就成了研究"quantum effects"，意思完全不

同。

## 1.11  形似词

(1)  "imagine"是"想象"，经常被错误地写成"image"（意思是"图像"）

(2)  "photo"是"照片"，"photon"是"光子"。

(3)  注意"assess"(评估、估计) 和"access"(访问、靠近、接近、取得) 的区别。

(4)  "complement (ary, arity) "(补充),"compensate"(补偿), "compliment(ary)" (赞扬)。

(5) "applying"不是"appling"(很多学生在以"y"结尾的单词加"ing"的时候都把"y"去掉)，同样的例子有：studying, flying, accompanying。实际上加"ed"的时候才是需要把"y"换成"i"：applied, studied, flied, accompanied.

(6) "accept"和"except"[5]
"accept"：接收、接纳；Please accept my gift.
"except"：  除……以外；We go to school everyday except Saturday and Sunday.
(7) "advice"和"advise"[5]
"advice"：忠告、劝告（名词）；She gives good advice.
"advise"：提出忠告（及物动词）；Please advise me on what to do.
(8) "affect"和"effect"[5]
"affect"：给予……影响；Do not let your personal problems affect the quality of your work.
"effect"：结果；The loss did not have an effect on me.
"effect"也有作动词的时候,意思是"提出"：She effected policies that benefited the entire organization.
(9) "adapt"和"adopt"[2]
"adapt"：适应（新环境、新学校等）；We were finally able to adapt to the cold climate of the area.
"adopt"：采纳（建议、措施）；The group adopted the strategy and implemented it in the entire unit.
(10) "latter"（指次序的后）和"later"（指时间的后）：

[1] We have studied the accretion disk models for both black hole X-ray binaries and active galactic nuclei; the **latter** (如果换成"later"就是错误的) harbor supermassive black holes with masses from millions to billions solar masses.

[2] We observed a neutron star first and a black hole **later** (如果换成"latter"就是错误的).

(11)  不能区分"it's"和"its"

[1] From equation (10), we can calculate **it's** (需要改成"its"）surface temperature accurately.

[1] From these data, we are confident that **its** (应该改成"**it's**"或者"**it is**",但是在科技论文中我们应该使用"**it is**"，而不是口语化的"**it's**"）a black hole.

**1.12    词组搭配**

(1) "spherically symmetric"和"spherical symmetry"是正确的。"spherical symmetric"和 "spherically symmetry"是错误的。

(2) "detail description"，"in details"是错误的，正确的是"detailed description"，"in detail".

(3) 介词后不能接介词，要接名词。"**Unlike in the case with**（改为"**Unlike the case of** ）the uniformity of the universe, no apparent violation of this law is known."

(4) 关于"take into account"的正确例子：

[1] We will take the force into account.

[2] We will take into account the force.

[3] We will take account of the force.

而"We will **take into account of** the force."是错误的。

(5) "These observations have made the studies of the formation processes of stars, planets, galaxies and quasars **possible**."需要修改为

"These observations have made **possible** the studies of the formation processes of stars, planets, galaxies and quasars."(请注意"**possible**"的位置)。

(6) "These black holes have masses **between $10^5$ — $10^{10}$** （修改为"**between $10^5$ and $10^{10}$**"）solar masses".

(7) "Tremendous observational evidence supporting the existence of black holes in the universe is gradually permitting the **uncovering** (应改为"**uncovering of**") the mysteries of black holes."（注意这里的"uncovering"是动名词，所以后面需要"of"）

(8) "The left panel **in** (应改为"**of**") Figure 9 shows a theoretical calculation of ... "

(9) "However this possibility, if true, may have fundamental impacts **to**（应改为"**regarding**"）the evolution and fate of the universe, as I will discuss **in** （应改为"**at**"）the end of this chapter."

(10) "compare to"用于表示"相似性"，"compare with"用于表示"差异性"。如：

[1] **Compared to** source B, source A shows a similarly hard spectrum. （指两个源类似）

[2] **Compared with** source B, source A shows a much harder spectrum. （指两个源不同）

(11) **Over than**（改为"**More than**"或者"**Over**"）20 experts at ESA side will **share us with**（改为"**share with us**"）their experiences regarding to how to build the ground segment and how to carry out the calibration."

(12)"consist in"表示"存在于…中"，"consist of"是"包括、由…组成"的意思

[1] A proton and an electron **consist in** a hydrogen atom.

[2] A hydrogen atom **consists of** a proton and an electron.

[3] A hydrogen atom **is consisted of** a proton and an electron.

(13) "These results **are well agree with** Zhang et al. (1999)" 改为

[1] These results **are in good agreement with** Zhang et al. (1999).或者

[2] These results **agree well with** Zhang et al. (1999).

(14) Our result is **in general** consistent with observations. 改为 Our result is **generally** consistent with observations.

<p style="text-align:center">表6 常用介词短语[2]</p>

| | |
|---|---|
| agree on "terms" | look over "an account" |
| agree to "a proposal" | proceed to "do something" (开始进行) |
| agree with "a person" | proceed with "doing something"(继续进行) |
| begin by "doing something" | prohibit from "doing something" |
| begin from "a point" | provide against "something" |
| begin with "an act" | provide for "something" |
| capable of "doing something" | provide one's self with "something" |
| capacity to "do something" | prefer one to the other |
| conform to/adapt one's self to | prefer to do one thing rather than another |
| conform with/in harmony with | preference for |
| consistent with | prevent from "doing something" |
| content one's self with | pursuant to/in pursuance of |
| correspond to/resemble | range from X to Y |
| correspond with/communicate with | reference to |
| differ with somebody | relief to suffering |

| differ from something | relieve one from a duty |
| evidence for a theory | accept responsibility for an action |
| evidence of something | responsibility to someone |
| indifferent to | result from something（源于某事） |
| join in a project | result in failure（导致失败） |
| look for a missing article | result of an investigation（调查结果） |

## 1.13    及物动词、不及物动词和情态动词的用法

(1) "cannot"是美国式用法（比如如果投稿到ApJ），"can not"是英国式用法（比如如果投稿到MNRAS）。

(2) "We also **referred** (应改为"referred to") Figure 3 and Figure 4 in the text."

(3) "However, in science direct evidence is not always what **leads** (应改为"leads to") the discovery of something." 注意"The appearance of X-rays **leads** the optical burst."（"X射线的出现早于光学爆发"）和"The appearance of X-rays **leads to** the optical burst."（"X射线的出现导致了光学爆发"）语法上都是对的，但是意思不一样，前一个是纯粹的时间先后，后一个是因果关系。当然"Prof. Zhang leads a very active research group."这里的"lead"就是领导的意思。

(4) "may, maybe, may be". 如：

[1] "This **maybe** the reason."应该修改为："This **may be** the reason."或者"**Maybe** this is the reason."

[2] (a) He agreed that this burst **may totally from** the external shock. 应该修改为：

[2] (b) He agreed that this burst **may come totally from** the external shock. 或者

[2] (c) He agreed that this burst **may originate totally from** the external shock. 或者

[2] (d) He agreed that this burst **may be totally from** the external shock.

但[2] (b) 中"**may**"如果换成"**maybe**"则是错误的。

## 1.14    "or"和"and"

"黑洞按照质量可以分为恒星级质量、中等质量和超大质量黑洞"，经常被错误地翻译成

"Black holes, according to their masses, can **divide** into stellar mass, intermediate mass **and** supermassive black holes."应该改为

"Black holes, according to their masses, can **be divided**（下文4.1节对语态有详述）into stellar mass, intermediate mass **or** supermassive black holes."

错误的原因：英文里面"**or**"和"**and**"的意思非常不同，"**or**"是指"非此即彼"，"**and**"是指"同时"。而中文的"和"的意思是比较模糊的，可以指"同时"也可以指"非此即彼"，所以经常会把中文的"和"直接就翻译成了"**and**"，其实很多情况下指的是"**or**"。但是中文的"或者"的意思则是比较明确的，但是使用不多。

## 1.15    用词不简洁或不够准确

表7    一些常出现的不简洁的用词及正确表示[2]

| 错误 | 准确 |
| --- | --- |
| research work | research |
| knowledge memory | memory |
| simulation results | simulation |
| knowledge information | information |
| calculation results | calculation |
| application results | application |
| two and a half decades | 25 years |
| 500 students in college | 500 college students |

(1) "The two scenarios **are almost equally well to** represent the high-z LGRB rate excess."可以修改为"The two scenarios **almost equally well** represent the high-z LGRB rate excess."

(2) It is found that although this case can well reproduce the observed log$N$—log $P$ distribution, but **much over produces** (改为"**significantly over-produces"**) the observed GRBs at $z \sim 2$ and **lack to produce** (改为"**under-produces"**) GRBs at $z > 3$.

(3) "**Despite of** (改为"**Despite"，属于词组搭配错误，见1.12**) tremendous progress in black hole research, many fundamental **questions concerning** (改为"**characteristics of"**) astrophysical black holes in the physical universe remain not fully understood or clarified."

(4) In this **paper**（改为"**study"**）, IDEAS was used to ….

(5) 美式拼写和英式拼写

有的单词在拼写上有英美两种不同的拼写，则给美国的期刊投稿时一定要采用美式拼写，

给欧洲期刊（尤其是英国的期刊）投稿时，采用英式拼写。常用的几类拼写差别见下表。

表 8 美式拼写和英式拼写[2]

|  | 美式拼写 | 英式拼写 |
| --- | --- | --- |
| 1） | -er | -re |
|  | center | centre |
|  | fiber | fibre |
|  | goiter | goitre |
|  | liter | litre |
|  | maneuver | manoeuvre |
|  | meter | metre |
|  | somber | sombre |
|  | theater | thertre |
| 2) | -ize/-yze | -ise/yse |
|  | organize | organise |
|  | realize | realise |
|  | analyze | analyse |
|  | catalyze | catalyse |
|  | modern ize | modernise |
|  | apologize | apologise |
|  | civilize | civilise |
|  | rationalize | rationalise |
|  | popularize | popularise |
| 3) | -log | -logue |
|  | analog | analogue |
|  | catlog | catalogure |
|  | dialog | dialoggue |
| 4) | -o | -ou |
|  | armor | armour |
|  | behavior | behaviour |
|  | clamor | clamour |
|  | color | colour |
|  | favor | favour |
|  | flavor | flavour |
|  | humor | humour |
|  | labor | labour |
|  | odor | odour |
|  | vigor | vigour |

# 2  标点符号

## 2.1 方程后面不加任何标点符号

由于方程是句子的一部分，前后(主要是方程后面) 都必须有标点符号。例如：

In Paper I and II, we constructed a phenomenological model for the dipole magnetic field evolution of pulsars with a long-term decay modulated by short-term oscillations,

$$B_{\mathrm{NS}}(t) = B_{\mathrm{d}}(t)(1 + \sum k_i \sin(\phi_i + 2\pi \frac{t}{T_i})), \tag{1}$$

where $t$ is the pulsar's age, and $k_i$, $\phi_i$, $T_i$ are the amplitude, phase and period of the $i$-th oscillating component, respectively. $B_{\mathrm{d}}(t) = B_0(t/t_0)^{-\alpha}$, in which $B_0$ is the field strength at the age $t_0$, and $\alpha$ is the power law index.

请注意：（1）方程后面有逗号"，"；（2）下标"NS"和"d"是正体，因为是单词的缩写，不是物理量或，其他的下标是斜体，因为代表变量；（3）数学函数"sin"是正体。

## 2.2 分号";"的使用

(1) 为了避免一个句子太长，或者强调主要的内容，通常英文中分号后面是对前面的补充（中文的分号是排比）。例如：

[1] I teach a graduate course on frontiers of astrophysics between 7:20 pm to 9:45 pm every Thursday**;** I deliberately arranged to have my classes in the evenings because ...

[2] In this work, a simple method is provided to combine GRB data into the joint observational data analysis to constrain cosmological models**;** in this method those SNe Ia data points used for calibrating the GRB data are not used to avoid any correlation between them.

(2) 分号";"也常用于分隔的含有逗号的并列成分[2]。

[1] We thank Gang Li, Tsinghua University, for timing data**;** Hong Zhao, Peking University, for helpful discussions**;** and the National Natural Science Foundation of China, for financial support.

[2] Follow this procedure: first, get your application forms**;** next, fill them out**;**
last, pay the charge.

(3) 经常看到学生不会在latex 文件里面正确输入引号。正确的做法：`test'生成ps或者PDF文件之后是'test'. 但是大部分学生都是输入：'test'. 类似的情况是：``test"而不是"test".

## 2.3 短破折号"en dash"的使用

也就是数学的"负号"（latex里面用 "--"，注意是两个"-"连着，或者$-$。word如果在文本模式下需要使用"插入"功能，选择里面的"负号"）， 相当于英文大写字母"N"的宽度，为连字符长度的2倍。代表"从……到……"，前后不空格。例如： 9 am$-$5 pm；1921$-$1949；chapters 8$-$9；Figure 1$-$4；5$-$50 kg；compounds A$-$I.

但数字有负号等符号修饰时，需要用"to"或"through"，不可用"短破折号"，如：

2 to +12 km；－1 to 3 mag；10 to >90 mL；<5 to 15 g等。

天体源的名字里面通常都有短破折号，但是通常都被错误地写成了连字符（见下面），比如 GRO J1655－40，通常被错误地写成了GRO J1655-40；Cyg X－1，通常被错误地写成了Cyg X-1.

## 2.4 连字符或连接号[2]

**2.4.1** 用于同时有多个前缀的复合词。如："mid-infrared"，"post-reorganization"，"bi-univalent"等。

**2.4.2** 用于含有"like"，"wide"等后缀的多音节词或已含有连字符复合词。如：resonance-like；radical-like；university-like；rare-earth-like；transition-metal-like 等。

**2.4.3** 用于复合词。在两个或多个单词组成复合词做修饰语的情况下，一般需要使用连字符以避免误解，如 American-football player。如果没有连字符，那么这个词组也可以理解为'美国籍的足球运动员（American football-player）'，而不是本来要表达的'美式橄榄球运动员'。这种复合词修饰语可以是更多个单词（如 ice-cream-flavoured candy），也可以做副词（如 spine-tinglingly frightening）。

**2.4.4** 用于由数字、字母或元素符号与名词或形容词组成的复合性修饰语。如：20th-century development；thirty-day period（注意常犯的错误是：thirty days period）；three-dimensional model；three-stage sampler；a 3-year-old child（注意常犯的错误是：a 3 years old child）；4-mm-thick layer 等。

[1] He is a **3 years old**（改为"3-year-old"）child.

[2] He is **3-year-old**（改为"**3 years old**"）.

**2.4.5** 用于由单一字母、元素符号或数字与名词或形容词组成的修饰语。如：K-Ar age；O-ring；X-band；x-axis；α-helix；γ-ray；π-electron 等。

**2.5 圆括号的使用**

**2.5.1** 用于句中表示补充信息，起解释、说明作用。如：

[1] The final step **(washing)** also was performed under a hood.

[2] The results **(Table 1)** were consistently positive.

**2.5.2** 括出表示序号作用的数字或字母。如：

(1), (2), (3), (a), (b), (c)。

**2.5.3** 用于数学公式或化学式。如：

$(k-1)/(g-2)$；$K_4Fe(CN)_6$ 等。

# 3 规范化

## 3.1 大小写

(1) SI 词头和有些单位的大小写容易出错

如能量的单位：eV (不是 ev)，keV (不是 KeV), MeV (不是 meV)，GeV，TeV。

表 8  SI 词头[7]

| 所表示的因数 | 中文名称 | 英文名称 | 符号 |
|---|---|---|---|
| $10^{24}$ | 尧[它] | yotta | Y |
| $10^{21}$ | 泽[它] | zetta | Z |
| $10^{18}$ | 艾[可萨] | exa | E |
| $10^{15}$ | 拍[它] | peta | P |
| $10^{12}$ | 太[拉] | tera | T |
| $10^{9}$ | 吉[咖] | giga | G |
| $10^{6}$ | 兆 | mega | M |
| $10^{3}$ | 千 | kilo | k |
| $10^{2}$ | 百 | hecto | h |
| $10^{1}$ | 十 | deka | da |
| $10^{-1}$ | 分 | deci | d |
| $10^{-2}$ | 厘 | centi | c |
| $10^{-3}$ | 毫 | milli | m |
| $10^{-6}$ | 微 | micro | μ |
| $10^{-9}$ | 纳[诺] | nano | n |
| $10^{-12}$ | 皮[可] | pico | p |
| $10^{-15}$ | 飞[母托] | Femto | f |
| $10^{-18}$ | 阿[托] | Atto | a |
| $10^{-21}$ | 仄[普托] | Zepto | z |
| $10^{-24}$ | 幺[科托] | yocto | y |

表9 常用的厘米克秒制单位

| 量的名称 | 单位 | 单位符号 | 定义 |
|---|---|---|---|
| 长度 | 厘米 | cm | 1 cm |
| | 秒差距 | pc | 1 pc=3.086×$10^{18}$ cm |
| 质量 | 克 | g | 1 g |
| 时间 | 秒 | s | 1 s |
| 力 | 达因 | dyn | 1 dyne = 1 g cm/s ² |
| 能量 | 尔格 | erg | 1 erg = 1 g cm ²/s ² |
| | 电子伏特 | eV | 1 eV = 1.60×$10^{-12}$ erg |
| 功率 | 尔格/秒 | erg/s | 1 erg/s = 1 g cm ²/s ³ |

(2) 特指的时代（或时期）、专有名词（人名、地名等）、星系、星座等名称首字母大写；与人名、地名等构成的名词首字母大写，但后面的普通名词不需大写。

the Stone Age; the Paleozoic Era; Glacial Epoch; Albert Einstein; the Milky Way; Venus Aries; Air force One; Titanic; Shenzhou spacecraft等。

(3) 方程后面如果不是句号，后面的紧接的句子不能另起一段，而且第一个字母必须是小写；方程后面如果是句号，后面紧接的句子可以另起一段，也可以不另起一段，但是第一个字母必须是大写。

(4) 冒号后面如果接的是句子，则第一个字母最好是大写（如果写成小写也不是不可以，有时候在正式文献也看到这么写，但是我们认为不够规范）；如果是单词或词组，则第一个字母是小写。

[1] It requires that a clock still show proper time after being read: The quantum uncertainty in its

position must not introduce significant inaccuracies in its measurement of time over the total running time. (请注意这里的"show"用动词原形，因为"require"是命令语气。)

[2] I classify black holes into three categories: mathematical black holes, physical black holes or astrophysical black holes. (这里用"or"，参见1.14节)

(5) 双引号内如果是完整的句子，第一个字母应该用大写：Some of the discussions, especially on the question "**will** (改为**Will**) all matter in the universe eventually fall into black holes?"**,** (**此处逗号应删除**) are quite speculative.

## 3.2 数词

(1) 如果能够使用一个单词的数一般不用阿拉伯数字，但是后面接物理量单位时，则用阿拉伯数字。

[1] Up to now, about 20 (改为twenty) black holes with masses around ten (应改为10) solar masses, called stellar mass black holes, have been identified observationally.

[2] All **3** (改为**three**) studies concluded that the mean temperature should be **30** ℃.

(2) 位于句首时，数字必须用英文单词不用阿拉伯数字表示。
[1]  **Twelve**（不用 **12**）parameters were selected for the experiment.
[2] Five hundred（不用 500） asteroids were detected using radar observations.

(3) 处于并列关系时，数字的表达形式不能混用。

Group of **eight** (改为**8**), 52, and 256 particles...;

Nine out of **752** samples.... (按照习惯**752**应该改为"**seven hundred and fifty-two**"，但是由于数字比较长这样不够简洁，所以当数字比较长的时候在科技文章里面习惯上也可以直接用数字)

(4) 2个或多个具有修饰关系的数字连用时，通常需要给出1个数字的英文单词形式，以免混淆。如 twenty 2 keV particles（20个具有2 keV能量的粒子）。

(5) 不确切的数目需要采用英文单词全拼形式。如

　　　the mid-sixties; the seventies等。

(6) 分数单独出现时需采用英文单词形式，分数不可与SI单位连用，且分数前面不可加不定

冠词**a**或**an**。如 one-half inch; the major axis is 0.25 (不是1/4) cm; two third of the sample was lost等。

(7) 连续的数字用阿拉伯数字表示，并采用to或en dash连接。如

1 to 2 m；140–150 m；5% to 15%；5%–15%等。

(8) 日期的写法：the first of July, 2006；July first, 2006；the second of July, 2006；the third of July；the 4th of July；the 28th of July；July 28th, 2006；2006-07-28；07-28-2006 (USA), 28-07-2006 (Europe and the rest of world)。

(9) 年代的表示：20世纪30 年代，"1930s"是正确的，"1930's"是错误的。

### 3.3 缩写

(1) "Professor"的缩写是"Prof."不是"Pro."，如"Dear Pro.Zhang"是错误的应该是"Prof. Zhang"（注意"."后面的空格）。"Doctor"的缩写是"Dr."不是"Doc."

(2) Figure 和 Table 的正确缩写分别是 Fig.和 Tbl.。缩写不能出现在句首，并且 Figure/Table 和数字之间必须有一个空格。

Figure.6, Figure6, Fig.6, Tbl10是错误的；Figure 6, Fig. 6, Tbl. 10是正确的。

### 3.4 正斜体

关于正体和斜体：正文里面数学符号、物理量符号等(如$x, y, z, A, B, C$这些数学符号都是能够取值的) 应该是斜体，单位是正体而不是斜体(如kg, km, eV)，各种函数名字，如sin, cos, log, ln 等应该是正体（latex的命令为\sin, \cos, \log, \ln，或者{\rm log}也可以；公式里面的文字说明(因为不能取值) 必须是正体(latex 中"\rm "命令的功能就是把后面的内容变成正体)。下标里面也是一样的规则。比如"黑洞质量"$M_{\rm BH}$ 是正确的，但是M$_{\rm BH}$、$M_{BH}$、M$_{BH}$ 都是错误的，因为"BH"不能取值，只能以正体出现。请注意当"$M_{\rm BH}$"作为单位在正文中使用的时候，"$M$"也是以斜体方式出现。类似地，"$M_k$" 是正确的，但是M$_k$、$M_k$、M$_k$ 都是错误的，因为"$M$"和"$k$"都可以取值，所以是斜体。

**3.5 留空格和不留空格的地方**

由于中文没有空格的概念，空格的使用错误在中国学生中极为普遍。

(1) 单位和前面的数字之间必须留空格，如10 eV, 20 kg, 40 km；在公式里面留空格$m=10\{\rm kg\}$，或者$m={\rm10\ kg\}$，或者$m=10\mbox{ } {\rm kg\}$（注意"\"和"\mbox{ }"的效果是一样的）。但是如果写成$m=10{\rm kg\}$，或者$m={\rm 10 kg\}$，则单位和前面的数字之间就不留空格。但是注意"2″，10′，5°，30°C"数字和单位之间不留空格。

(2) 左括号"("和前面的字母之间必须留空格，和后面的字母之间不留空格；右括号")"和前面的字母之间不留空格，和后面的字母之间留空格，右括号")"的后面是标点符号则无空格。逗号 "," 和点号"."和前面的字母之间不留空格，和后面的字母之间留空格。

[1]下面这个句子里面有7处空格使用错误的地方，请读者留意："This problem has been solved(Einstein 1915 )and consequently applied to many astrophysical settings( Freedman 1932**;**Lieu 1945) ."

改为 This problem has been solved (Einstein 191**5**) and consequently applied to many astrophysical settings (**F**reedman 1932**;** Lieu 1945).（注意黑体标出的地方"空格"的使用）。

[2] From Figure **3,** we find that Dr**.** Liu's calculations…（注意黑体标出的地方"空格"的使用）非常普遍的错误是："From Figure**3,**we find that Dr.Liu's calculations…"，或者"From Fig.**3,**we find that Dr.Liu's calculations…"（注意"Fig.3,we"应该改为"Fig. 3, we"，也就是需要在"."和","后面都需要留空格。"Dr.Liu"应该改为"Dr. Liu"。）

(3) 用上标表示的单位出现小数点的情况：0.1角秒(0″.1)，0.1角分(0′.1)，0.1度(0°.1)，1.1角秒(1″.1)，1.1角分(1′.1)，1.1度(1°.1)，这里单位和数字之间没有空格。

(4) "/"前后一般都没有空格，比如：The anonymous referee is thanked for **his / her** （须改为**"his/her"，也就是不要留空格**）comments and suggestions.

**3.6 参考文献的引用方法**

(1) 英文期刊参考文献在句子中充当主语或宾语：

[1] Zhang et al. (1999) have shown that black holes can eat matter...

[2] It has been shown by Zhang et al. (1999) that black holes eat matter.

(2) 英文期刊参考文献在句子中不充当任何成分：

[3] It has been known that black holes can eat matter (Zhang et al. 1999).

常见的错误例句是：

[1] Zhang et al. 1999 have shown...

[2] It has been shown by Zhang et al. 1999 that black holes eat matter.

[3] It has been known that black holes can eat matter (Zhang et al. (1999)). （请注意括号的位置。）

### 3.7 制图

**3.7.1** 坐标轴需要名称和单位。

**3.7.2** 几乎所有绘图程序缺省设置的图上的字母和数字字体都太小，基本原则是按照杂志要求的尺寸印出来后图上的字母和数字和正文的字体相当或者略大，但是应该比"figure caption"的字体大一些。

**3.7.3** 关于图的一些英文说明：

(1) 一个figure 如果由几个图组成，每个图叫做panel，而不是sub-figure，图的位置通常称为top panel, middle panel, bottom panel, upper panel, upper-left panel, lower-right panel, 等等；

(2) 图中线条的说法是solid line, dotted line (不是dot line), dashed line (不是dash line), dot-dashed line。

## 4 语法分析

### 4.1 语态

(1) 尽量不要使用主动式，特别是"People have done... "，应该写"It has been calculated... by xxx"(如果你不想强调到底是谁做的，而仅仅想列出是谁做的，就不要写"by xxx"，而只要列出参考文献就可以了)。

(2) 应该使用被动语态的却使用主动语态

[1] "Several issues about black hole growth **have clarified**."应该修改为"Several issues about black hole growth **have been clarified**."

[2] "The configuration is consists of three parts."是错误的。有两个正确的写法：

(a)　The configuration is consisted of three parts.

(b) The configuration consists of three parts.

[3] "模型分为三类"经常被写成"Models **divide** into three classes"，正确的写法是"Models **are divided** into three classes"。

[4] 黑洞位于星系的中心："Black hole **locates** in center of galaxy."是错误的，应该改为"A black hole **is located** in the center of a galaxy."

每一个星系的中心都有一个黑洞："A black hole **is found** in the center of every galaxy."

或者"Every galaxy **harbors** a black hole at its center."

所观测的5个星系中心都有一个黑洞："Each of the five galaxies observed **harbors** a black hole at its center."或者"A black hole **is found** (located) in the center of each of the five galaxies observed."

[5] "the gravitational potential energy **transforms to** the radiation on NS surface."改为"the gravitational potential energy **is converted** to the radiation on NS surface."

[7] "The observed spectral variation may attribute to the variation of the electron energy distribution."(意思是观测到的谱变化可能是由电子能量分布的变化引起的)可以改为"The observed spectral variation may **be attributed to** the variation of the electron energy distribution."或者"The observed spectral variation may **be due to** the variation of the electron energy distribution." 或者"The variation of the electron energy distribution may **attribute to** the observed spectral variation."

[8] The bubble **nebula result** from the radiatively driven outflow during the different X-ray active phases. 两个错误：（1）nebula 是单数（复数是 nebulae），所以后面的动词应该用第三人称单数形式；（2）但是主要的错误是"bubble nebula"是"被"产生出来的，所以应该使用被动式。所以正确的句子应该是 The **bubble nebula is resulted** from the radiatively driven outflow during the different X-ray active phases. 如果用主动式，可以这么写：The outflow **results in** the bubble nebula. 注意"is resulted from"和"results in"的用法。

[9] "The gas pressure **relates to** （或 **associates with，都是错误的**） its temperature."改为"The gas pressure **is related to** its temperature."

(3) 应该是主动语态的地方被误用成被动语态

[1] After the correction, the peak in the spectrum **is disappeared**. (改成 "has disappeared" 或者 "is absent").

[2] "All models **are included** an interstellar absorption component with the hydrogen column…" 改为"All models **include** an interstellar absorption component with the hydrogen column…"

## 4.2 时态

关于时态的正确使用：

(1) 叙述你本论文的工作使用一般现在时；

(2) 描述你以前的相关工作使用一般过去时；

(3) 描述别人以前的工作使用一般过去时；

(4) 但是当你现在的工作是以前你的或者别人的某个工作的延续时，对那个工作的描述需要使用一般完成时；

(5) 在文章的summary部分描述本文章的结果时可以使用一般现在时、过去时和一般完成时，如"The black hole we **studied** in this work ... ","In this paper we **have calculated**... ","It **has been shown** that ... ","We **study** this problem ... ","We **analyze** these data... ","We **start** from ... ","We found that ... ","We **conclude** that","It **is shown**","In this paper we **calculate** ... ";

(6) 在回复审稿人的信里面，谈到你这篇文章中你做了什么的时候用一般过去式，或者一般完成时，如：

[1] Thanks for the referee's suggestion. Actually we **calculated** the mass accretion of three sources, but neglected to include the result in the paper.

[2] We **included** Figure 2 and Table 3 in the revised manuscript.

[3] We **thank** the referee for these insightful comments and suggestions.

[4] We **have improved** the manuscript substantially.

[5] The manuscript **has been improved** significantly by including...

[6] 如果是回复审稿人的第二封信，开头可写为：We **have further improved** the manuscript following the suggestions in the second referee report...

(7) 描述发生在过去的一件事用过去时，描述虽然过去发生的、但是已经被认定为"普遍事

实(general truth)"的使用一般现在时，如：

[1] Homan et al. (2006a, 2006b, 2007a) **suggested** that XTE J1701-462 **was** a Z source, because **it exhibited** typical CDs of Z sources and its timing properties **were** also consistent with those of Z sources.

例句[1] 并没有错误，但是如果写成：Homan et al. (2006a, 2006b, 2007a) **suggested** (发生在过去) that XTE J1701-462 **is** (普遍事实) a Z source, because it **exhibited** (发生在过去) typical CDs of Z sources and its timing properties **are** (普遍事实) also consistent with those of Z sources. 则意义有所不同，更加强调"XTE J1701-462 is a Z source"和"its timing properties are also consistent with those of Z sources"是被认定的"普遍事实"。

再举两个例句：

[2] In yesterday's battle, three soldiers **were killed** and one **was injured**.

[3] Now a total of 15 soldiers **are dead** and nine soldiers **are injured**.

# 5 句法分析

## 5.1 句子成分的缺少或多余

(1) 宾语里面不要出现不特指的人，很多文章出现"This has enabled **people** (应删除people) to do"。

(2) "get"(包括"got") 使用的太多，可以适当使用"obtain", "derive", "deduce", "acquire"等(不要使用类似这样的句子："This result **is got**.")。

(4)  importantly 和  important

[1] Very **importantly**, SNe eject a lot of matter... "是正确的，因为"importantly"对应的是"eject"，所以需要使用副词。

[2] Very **importantly**, SNe are powerful fountains of high energy particles... "是错误的，因为"are"要对应形容词，所以应该修改为: Very **important**, SNe are ……

(5) "As one of our valued authors **I** believe you **may interested to** hear about IOP Asia-Pacific, a new website from IOP Publishing dedicated to showcasing research from the Asia-Pacific region."这是一个英文期刊的编辑部发给笔者的信里面的一句话，但是这句话有两个错误，

一个是主语不清楚，一个是被动用错了。需要修改为"**I believe you, as one of our valued authors, may be interested to** ... "。

## 5.2 respectively 在句子中的位置错误[2]

(1) Equations 2—6 can be **respectively**（在这里的意思就成了"**尊敬地**"，显然不是作者的原意，作者的原意应该是"**分别地**"）linearized as:……(equations given)…

正确例子：　Equations 2—6 can be linearized as:……(equations given)…, **respectively**.

(2) The weights of the two experts are **respectively** 0.600 and 0.400.

正确例子：　The weights of the two experts are 0.600 and 0.400, **respectively**.

## 5.3 句子开头

(1) "So"作为句子的开头太频繁，可以适当使用"Therefore", "Hence", "Consequently", "Thus", "As a result"等。因此应该避免使用"So"作为句子的开头。

(2) "And", "But"作为句子的开头太频繁，读起来很难受。应该尽量不用"And", "But"作为句子的开头，可以适当使用"However"，"Nevertheless"等作为句子的转折。

(3) How to 不能作为句子开头。

**How to (改为Determining how to)** find the optimal parameter is the main objective.

(4) "We"作为句子的开头太频繁：只有作者特别想强调是"我们"做的，而不是"别人"做的时候才使用"We"作为句子的开头，在一篇文章的开始和总结部分可以适当地使用一些，正文中间一般都是使用倒装句字。

(5) "Now"作为句子的开头太频繁。因为科技论文的基本时态就是一般现在时，所以基本上没有必要使用"Now"(见关于时态的讨论)。

(6) 尽量不要出现"We feel ... "或者"We think ... "之类的句子。比如"We **think** the accretion rate dominates the …."应该改为"We **consider that** the accretion rate dominates the …."

(7) 科技论文中不用下列口语化的用法，应该在科技论文中避免："it's", "can't", "couldn't","doesn't" 等应该改成"it is", "cannot(美式英语)或者can not(英式或者欧式英语)", "could not","does not"。

很多人都喜欢用"a little"，这是很口语化的用法，例子：

[1] Its surface temperature is **a little** low. 改为

[2] Its surface temperature is **slightly too** low. 或者

[3] Its surface temperature is **somewhat** low.

(8) 除了命令式之外，一个完整的句子必须有主语（而且只能有一个)、谓语和宾语。很多同学在一个句子里面用多个主语。如

[1] The black hole binary is a very exciting source of high frequency QPOs, we therefore decided to study it. 是错误的，因为这个句子有两个主语："the black hole binary"和"we"。正确的写法是：

[1] **We** decided to study the black hole binary, because it is a very exciting source of high frequency QPOs. 或者

[1] **The black hole binary** is a very exciting source of high frequency QPOs**; we** therefore decided to study it. （注意这里分号"；"的用法）

再给一个例句：

[2] The black hole **binaries contain** a primary black hole and a companion star, the mass exchange between them through an accretion disk **in** the well-known thin accretion disk model. 里面有几个错误，修改后如下，

[2] A black hole **binary contains** a primary black hole and a companion star, in which the mass exchange between them **is** through an accretion disk **described by** the well-known thin accretion disk model.

## 5.4 谓语使用不当或与主语不一致

(1) "It **is come** (或者coming) from ... "应该改为"It **comes** from"或者"It **is** from"（它来自于）；It **is originated** from 改为 It **originates** from（它起源于） .

(2) "This radio-quiet quasar **is also have** (或者**having**) a relativistic jet."改为"This radio-quiet quasar also **has** a relativistic jet."

(3) "There **will have** an ESAC-HXMT workshop in March 7—8 at Villanueva de la Canada, Spain."改为 "There **will be** an ESAC-HXMT workshop in March 7—8 at Villanueva de la Canada, Spain."

(4) "There **are exist** three black holes in this galaxy"或者"There **are** three black holes **exist** in this galaxy"都是错的。应该改为 "There **are** three black holes in this galaxy"或者"There **exist** three black holes in this galaxy" 或者"Three black holes **exist** in this galaxy" 或者"Three black holes **are** in this galaxy"，这些都是对的。

(5) The TeV **data** of this SED in Figure 1 have considered the correction of the EBL absorption. （"data"怎么会"考虑"？）。应该改为 The TeV **data** of this SED in Figure 1 **have included** the correction of the EBL absorption.或者 The correction of the EBL absorption for the TeV data of this SED in Figure 1 **has been considered**.

(6) "The gas pressure **grows** with the growth of its temperature grow."改为"The gas pressure **increases** with the **increase** of its temperature grow."（把"增加（increase）"和"增长（growth）"混淆了）。与此对应的是"decline"和"decrease"的混淆。

(7) This result **consistents** with our model.句中 consistent 是形容词,不是动词,应该改为"This result **is consistent** with our model."

  (8) equal 即是形容词又是动词

*A* **is equal to** *B*; *A* **equals** *B*.都是对的，但是 *A* **equals to** *B*.就是错误的。

### 5.5 复合句

(1)  复句里面"which"和"that"作为代词连接从句和主句时，其先行词是整个句子，所以其后谓语动词保持一致用第三人称单数。"that"分句对先行词起限制和定义的作用，用于限制性关系分句；"which"分句对先行词进行评注或解释，用于非限制性关系分句，通常用逗号分隔。由于在很多情况下比较难确定该使用"that"还是"which"，所以现在习惯上"that"和"which"是经常通用的。但是如果细细体会，还是能够区分它们用法的不同的。一个大体都适用的简单原则是，如果用逗号分隔，就用"which"，否则就用"that"。

[1] Thus the temperature is determined by the heating of the central source's luminosity, **which is** (注意这里需要"**is**") given by equation (2).

[2] This is the sample **that**（这里就不能用"which"）Jack collected.

[3] The sample, **which**（这里如果用"that"就不太顺）jack collected, was lost in transit.

[4] I am returning the reports, **which** I have read.（"归还的报告"我都看过了)

[5] I am returning the reports **that** I have read.（我仅仅把"我看过的报告"归还了）

(2) 一个句子作为宾语时，该句子前面应该有"that"等连接词。

We **find** (改为"**find that**") the ΛCDM model is consistent with the joint data in 1−σ confidence region.

(3) "which"引导非限定性定语从句，"that"引导限定性定语从句。

[1] It is actually the expansion of the event horizon **which** (改为**that**) swallows the infalling shell.

[2] Almost all gravitational potential energy is converted into the kinetic energy of the accreted matter **which** (改为"**that**") free-falls into the black hole and thus **is** (注意这里需要"**is**") lost into the event horizon of the black hole.

[3] I am thinking about the environment distribution **to** (改为**that may**) cause this missing emission.

[4] There **are** two XRT observations **deviated** (改为 that deviate) from the constant radius.

(4) 从句里面主谓颠倒

[1] It was not clear who did it and why **was it** (改为**it was**) necessary to make the universe suitable for our existence.

[2] We know **who is** the author. 改为 We know **who** the author **is**.

[3] We know **what is** the answer. 改为 We know **what** the answer **is**.

[4] We know **why is** the universe expanding. 改为 We know **why** the universe **is** expanding.

[5] We know **how does** it work. 改为 We know **how it** works.

(5)从句缺主语

[1] There **is** no previous work **can** rule out this model."改成：

" There is no previous work **that** can rule out this model."或者 "No previous work **can** rule out this model."

[2]"There **is still no mature theory can** explain all X-ray properties of super-critical accreting sources."应该改为"There **is still no mature theory that can** explain ..."

[3]"There **is a subclass of active galactic nuclei (AGNs) are** candidates for super-Eddington accretion."应该改为"There **is a subclass of active galactic nuclei (AGNs) that** are candidates for super-Eddington accretion."

(6) 句子表述不够简洁

"The fit parameters listed in Table 3 are more accurate than those listed in Table 5, because, as shown by the top panels in Figure 2, the count rate variation within each of the 29 HID regions used to group data for producing the PCA spectra is extremely minimized, while, as shown by the bottom panels in Figure 2, the count rate varies substantially within most of the eight segments used to group data for producing the broadband spectra."这是从一个中国学生的论文初稿里面直接摘出来的,中国式英语的味道极浓, 如果直接翻成中文就不会有问题, 但是作为英文句子, 主语很多, 结果超复杂! 可以简化为:

"The fit parameters listed in Table 3 are more accurate than those listed in Table 5, because the count rate variation is minimized within each of the 29 HID regions used to group data for producing the PCA spectra, as shown by the top panels in Figure 2. However, the count rate varies substantially within most of the eight segments used to group data for producing the broadband spectra, as shown by the bottom panels in Figure 2."

**5.6 要表达的主要含义放句子前面，目的、位置、原因和时间等放句子后面。**

[1] For the application in galaxy formation, this paper studies ...
改为：This paper studies … for application in application in galaxy formation.

[2] In practice, we employed this approach to …, and the results show that this method is feasible.
改为： We employed this approach to …, and the results show that this method is feasible. （"in practice"和"feasible"意思重复, 即使不重复, 最好也是放到句子的最后。）

[3] To ensure sufficient signal to noise ratio, this source was observed for three days.
改为：This source was observed for three days to ensure sufficient signal to noise ratio. （前一个表达并没有语法错误, 但是强调的是"to ensure"而不是"was observed for three days"。）

[4] When $f$ is taken as the control parameter, the inferred values of $g$ are shown in Fig. 8.
改为：The inferred values of $g$ are shown in Fig. 8 when $f$ is taken as the control parameter. （前一个表达并没有语法错误, 但是强调的是"When $f$ is taken…"而不是"the inferred values…"。）

[5] Based on the observations presented here, it is found that this system contains a black hole.
改为：It is found that this system contains a black hole, based on the observations presented here. （前一个表达并没有语法错误, 但是强调的是"Based on the…"而不是"it is found that…"。）

[6] Because this source is very luminous and shows extremely strong variability, we thus conclude it must contain a black hole.

改为：We thus conclude this source must contain a black hole, because it is very luminous and shows extremely strong variability.

最后再说明一下，中文的习惯是"因为…所以"，但是英文的习惯是"所以…因为"，这是造成不通顺（或者不能正确表达意图）的主要原因。

## 5.7 两个科技英语中常用的句型

(1) "A 随着B 增加而增加"的正确翻译是："*A* increases with **increasing *B*.**"或者"A increases with B."例如：

[1] However, when the source moves on the upper track, the inner disk radius increases **with increasing of** (改为"**with**") the disk accretion rate.

[2] Panel B in Figure 11 shows the inferred NS surface magnetic field strength during the episodes in which the inner disk radius increases with the **increasing of the** (改为"**increasing**") disk accretion rate.

错误的翻译：

[1] A increases by increasing B.

[2] A increases with B increases.

[3] A increases with increases B.

(2) "A 随着B 增加而减少"的正确翻译是："*A* decreases with increasing *B*."下面给出例句：

The inner disk radius decreases with **increasing of the** (改为"**increasing**") disk accretion rate.